\documentclass[aps,prb,twocolumn, showpacs,superscriptaddress,preprintnumbers,amsmath,amssymb,longbibliography]{revtex4-2}

\usepackage[usenames,dvipsnames]{xcolor}

\usepackage[normalem]{ulem}
 \newcommand{\stkout}[1]{\ifmmode\text{\sout{\ensuremath{#1}}}\else\sout{#1}\fi}

\usepackage{float}
\usepackage{graphicx}
\usepackage{dcolumn}
\usepackage{bm}
\usepackage{multirow}
\usepackage{amsmath,amssymb}

\begin{document}
\title{High-field ultrasound study of elastic constants and possible magnetic symmetry transformations in UO$_2$}

\author{E. A. Tereshina-Chitrova}
    \affiliation{Institute of Physics, Czech Academy of Sciences, 18121 Prague, Czech Republic}
 
 \author{L. V. Pourovskii}
    \affiliation{Centre de Physique Th\'eorique CNRS, \'Ecole polytechnique, Institut Polytechnique de Paris, 91120 Palaiseau, France}
     \affiliation{Coll\`ege de France, Université PSL, 11 place Marcelin Berthelot, 75005 Paris, France}
 
 \author{S. Khmelevskyi}
    \affiliation{Vienna Scientific Cluster Research Center, Technical University of Vienna, Operngasse 11, 1040 Vienna, Austria}
 
 \author{D. Gorbunov}
    \affiliation{Hochfeld-Magnetlabor Dresden (HLD-EMFL), Helmholtz-Zentrum Dresden-Rossendorf, Dresden 01328, Germany}
     
 \author{R. Caciuffo}
    \affiliation{Istituto Nazionale di Fisica Nucleare, Via Dodecaneso 33, IT-16146 Genova, Italy}

\date{\today}

\begin{abstract}

Abstract: In this study, we probe the coupling between magnetism and lattice dynamics in UO$_2$, a 3\textbf{k} antiferromagnet that undergoes magnetic ordering below its N\'eel temperature (T$_N$) of 30.8 K. Ultrasound measurements provide insights into the interplay between the material's magnetic properties and lattice vibrations in response to the applied high magnetic field. A model analysis based on\textit{ ab initio} calculated superexchange interactions predicts a change in the magnetic structure from 3\textbf{k} to 2\textbf{k} at around 50 T. Although this change is not evident in the magnetization curve, we observe a crossover of ultrasound velocity $\Delta v/v$ using a phase-sensitive pulse-echo technique in pulsed magnetic fields of up to 65 T. We show that at H$_z$ $>$ 50 T, the structure remains antiferromagnetic in the ($x, y$) plane and becomes ferromagnetic in the $z$ direction. A further transition into the 1\textbf{k} structure is theoretically predicted to take place at a magnetic field of $\sim$104 tesla.
\end{abstract}


\maketitle

\section{Introduction} \label{Intro}
Beyond its critical role in nuclear reactors \cite{wang20}, UO$_2$ is recognized as a complex magnetic material that poses significant challenges to both experimental and theoretical research \cite{lander20}. This complexity arises from the intricate interplay of spin-phonon interactions, multipolar ordering, and Jahn-Teller effects, not easily detectable using conventional experimental techniques. UO$_2$ is a Mott insulator, which undergoes antiferromagnetic ordering below the N\'eel temperature (T$_N$) of 30.8 K through a first-order transition. In the ordered state, a transverse 3\textbf{k} antiferromagnetic arrangement is established, with magnetic moments of $\mu_0$ = 1.74(2) $\mu_B$ per uranium atom \cite{blackburn05,ikushima01} pointing in the $\langle$111$\rangle$ directions. Experimental evidence \cite{faber75,wilkins06} has highlighted the significant role of quadrupolar interactions in stabilizing the ground state of UO$_2$. Indeed, long-range antiferro-ordering of electric-quadrupole moments at the uranium site and a static Jahn-Teller distortion of the oxygen cage occur along with the magnetic transition at T$_N$. Moreover, a dynamical Jahn-Teller effect associated with a spin-lattice quadrupolar coupling was also observed above T$_N$ \cite{amoretti89,caciuffo99}. Theoretical calculations have indicated that the anisotropy of quadrupole superexchange may be the underlying mechanism behind the noncollinear 3\textbf{k }antiferromagnetic order in UO$_2$ \cite{pourovskii19}. Additionally, UO$_2$ exhibits strong coupling between its magnetism and lattice vibrations, as demonstrated by early studies in the 1960s, which observed significant anomalies in its elastic behavior near T$_N$ and across a broad temperature range above it \cite{brandt67,brandt68}.

Changing external conditions, such as the application of a strong magnetic field, can redistribute the energies associated with electronic, multipolar, and collective dynamic interactions, ultimately leading to modifications in the material's properties, including its magnetic structure. The recent magnetostriction studies on UO$_2$ \cite{jaime17} have provided fascinating insights into the material's behavior under extreme conditions. The exceptionally strong piezomagnetic behavior (linear magneto-elastic memory effect) of UO$_2$ has been revealed due to magnetic domain switching occurring around $\pm$18 T at low temperatures. However, no signals indicating potential field-induced magnetic structure changes in UO$_2$ were observed, even in magnetic fields as strong as 150 T \cite{schoenemann21}. This suggests that the energy scale of competing magnetic anisotropy and exchange interactions in the material is exceptionally high, possibly requiring even stronger external fields to shift the magnetic structure.

However, a recent single-crystal x-ray diffraction study \cite{antonio21} conducted in magnetic fields up to 25 T and within the magnetically ordered state has revealed an interesting phenomenon. For the magnetic field applied along the easy $[111]$ direction, a complex magnetic-field-induced evolution of the microstructure was found, reflected in the splitting of a specific diffraction peak. Lander and Caciuffo \cite{lander21b} suggested that this is a manifestation of a rhombohedral distortion of the unit cell in the field. This implies that a sufficiently strong magnetic field can alleviate the frustration of the zero-field 3\textbf{k} magnetic moment arrangement, stabilizing alternative magnetic structures. However, this question remained unresolved to date.

In this work, we explore the critical phenomena associated with non-collinear magnetism and complex electronic configurations in UO$_2$ by employing a combination of high magnetic fields and ultrasound measurements - both powerful tools for investigating quantum effects in materials. Ultrasound is particularly effective at detecting subtle changes in properties related to the reorganization of magnetic moments and electric quadrupole moments, owing to the strong coupling between the elastic strain field generated by sound waves and the quadrupole moments \cite{luthi07}. To ensure high accuracy of the measurements we use a phase-sensitive detection technique in pulsed magnetic fields \cite{zherlitsyn14}. 

We also present high-field theoretical magnetization data for UO$_2$, obtained through first-principles calculations, and model possible changes of magnetic structure across various field strengths. The theoretical insights indicate a phase transition at a relatively moderate magnetic field of $\sim$50 T along the \textit{hard} magnetization axis of the UO$_2$ single crystal, a direction that has been largely overlooked in prior high-magnetic-field studies \cite{antonio21,schoenemann21,jaime17}. Ultrasound measurements reveal a crossover at this field strength, suggesting that the transition may involve a smooth rotation of the magnetic moments rather than abrupt changes. Notably, this crystal direction is the most "anisotropic" in terms of elastic properties \cite{brandt67} and thermal conductivity \cite{gofryk14} in UO$_2$. This work contributes to a broader understanding of the complex interactions in UO$_2$ by providing new perspectives on the material’s magnetic behavior under extreme conditions.

\section{Experimental details}

\begin{figure*}
\centering
\includegraphics[width=2.0\columnwidth]{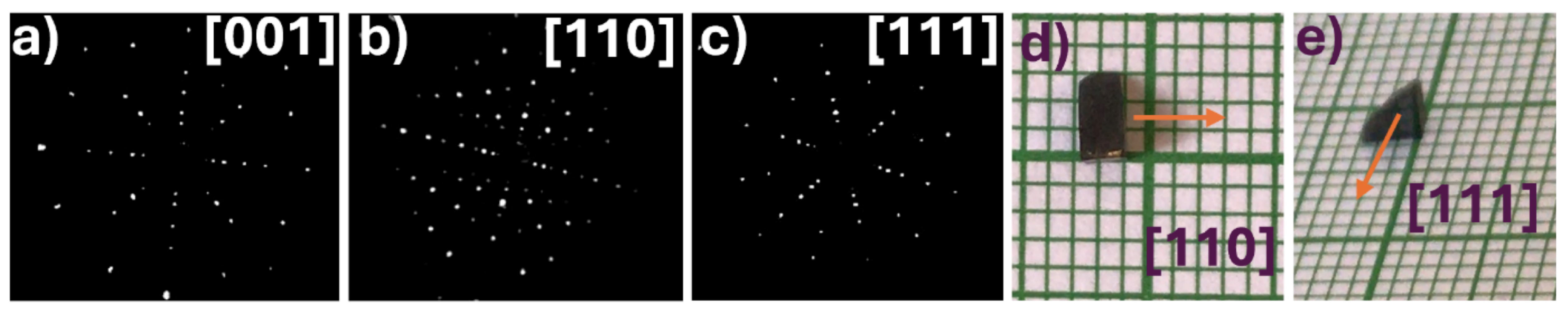}
\caption{(a-c) Back-scattered Laue patterns of a UO$_2$ single crystal cut in different directions, (d-e) Illustration of UO$_2$ samples for different measurement geometries (see text for details): (d) The parallelepiped-shaped sample has the [001] direction oriented out of the image plane. (e) For the triangular-shaped single crystal, the [111] direction is perpendicular to the triangular face. 
}
\label{Fig1}
\end{figure*}

A 380 mg single crystal of UO$_2$ with depleted uranium came from the Joint Research Centre (Karlsruhe) of the European Commission. We cut the crystal into three samples, two for ultrasound and one for magnetization measurements. The samples were oriented using back-scattered Laue patterns and cut and polished to produce surfaces aligned with the main crystallographic directions (Fig. \ref{Fig1}). Prior to measurements, the samples were annealed at 1200 $^{\circ}$C in an Ar atmosphere to minimize surface damage from the polishing and to ensure a bulk oxygen stoichiometry of UO$_2$.

The field and temperature dependencies of the relative sound-velocity, $\Delta v/v$, and sound-attenuation, $\alpha$, were measured using a phase-sensitive pulse-echo technique\cite{luthi07} in zero magnetic field, in steady field of 17 T, and in pulsed magnetic fields up to 65 T. A pair of piezoelectric transducers was glued to opposite surfaces of the sample cut in different crystallographic directions in order to excite and detect acoustic waves. The ultrasound frequencies varied between 28 and 33 MHz in our experiments. We measured the longitudinal $C_{11}$ (\textbf{k} $\|$ \textbf{u} $\|$ [001]), and transverse $C_{44}$ (\textbf{k} $\|$ [001], \textbf{u} $\|$ [110]) elastic moduli in magnetic fields along [001] and [111] directions as well as the properties along the easy magnetization directions $\langle$111$\rangle$, \textit{i.e.} \textbf{k} $\|$ \textbf{u} $\|$ \textbf{H}$\|$ [111]. Here, \textbf{k} and \textbf{u} are the wave vector and polarization of the acoustic waves, respectively. The $c$-axis of the magnetic structure in cubic UO$_2$ can align along any of the three cubic axes in a single crystal, and we will refer to the geometric measurement directions as indicated in Fig. \ref{Fig1}(d,e). The absolute values of sound velocity at 4.2 K for the measured acoustic modes are $v_{11}$ = 5200 $\pm$ 200 m/s and $v_{44}$ = 2350 $\pm$ 200 m/s. The sound velocity is related to the elastic modulus $C_{ii}$, via the relation $C_{ii}$ = $\rho v^2_{ii}$, where $\rho$ = 10.97 $\times$ 10$^3$ kg/m$^3$ is the mass density of UO$_2$. Using this relation, the elastic properties of UO$_2$ can be determined.

High-field magnetization was measured in pulsed magnetic fields up to 58 at the Dresden High Magnetic Field Laboratory \cite{zherlitsyn12}. Absolute values of magnetization were calibrated using data obtained in static fields up to 7 T using a SQUID magnetometer (Quantum Design, USA).

\section{Results and discussion} \label{results}
\subsection{\textit{High-field experiments}}

To provide essential context for understanding the magnetic phase transitions in UO$_2$, we first present the temperature dependence of the magnetic susceptibility ($M/H$) of a UO$_2$ single crystal, measured along the easy magnetization direction after field cooling in zero magnetic field (ZFC) (Fig. 2). A sharp first-order-like anomaly is observed at approximately 31 K, marking the onset of the antiferromagnetic order. Since the earliest studies of elastic properties of UO$_2$ \cite{brandt67,brandt68}, it has been evident that all elastic constants ($C_{11}$, $C_{12}$, $C_{11}$-$C_{12}$, $C_{44}$ for a cubic crystal) undergo significant changes at T$_N$. Among these, $C_{44}$ shows the most pronounced anomaly, including a discontinuity in attenuation at T$_N$ \cite{brandt67}. This behavior arises from a simultaneous magnetic transition and electric quadrupolar 3\textbf{k} AF ordering, accompanied by a static Jahn-Teller distortion of the oxygen cage \cite{wilkins06, lander20}, while the unit cell remains cubic, albeit at a reduced volume \cite{brandt67}. The current study indicates a pronounced 12\% softening of the transverse sound velocity at T$_N$, reflecting the strong magnetoelastic coupling in UO$_2$, a feature commonly observed in actinide compounds where magnetism interacts significantly with lattice dynamics \cite{lander74}. 

\begin{figure}
\centering
\includegraphics[width=1.0\columnwidth]{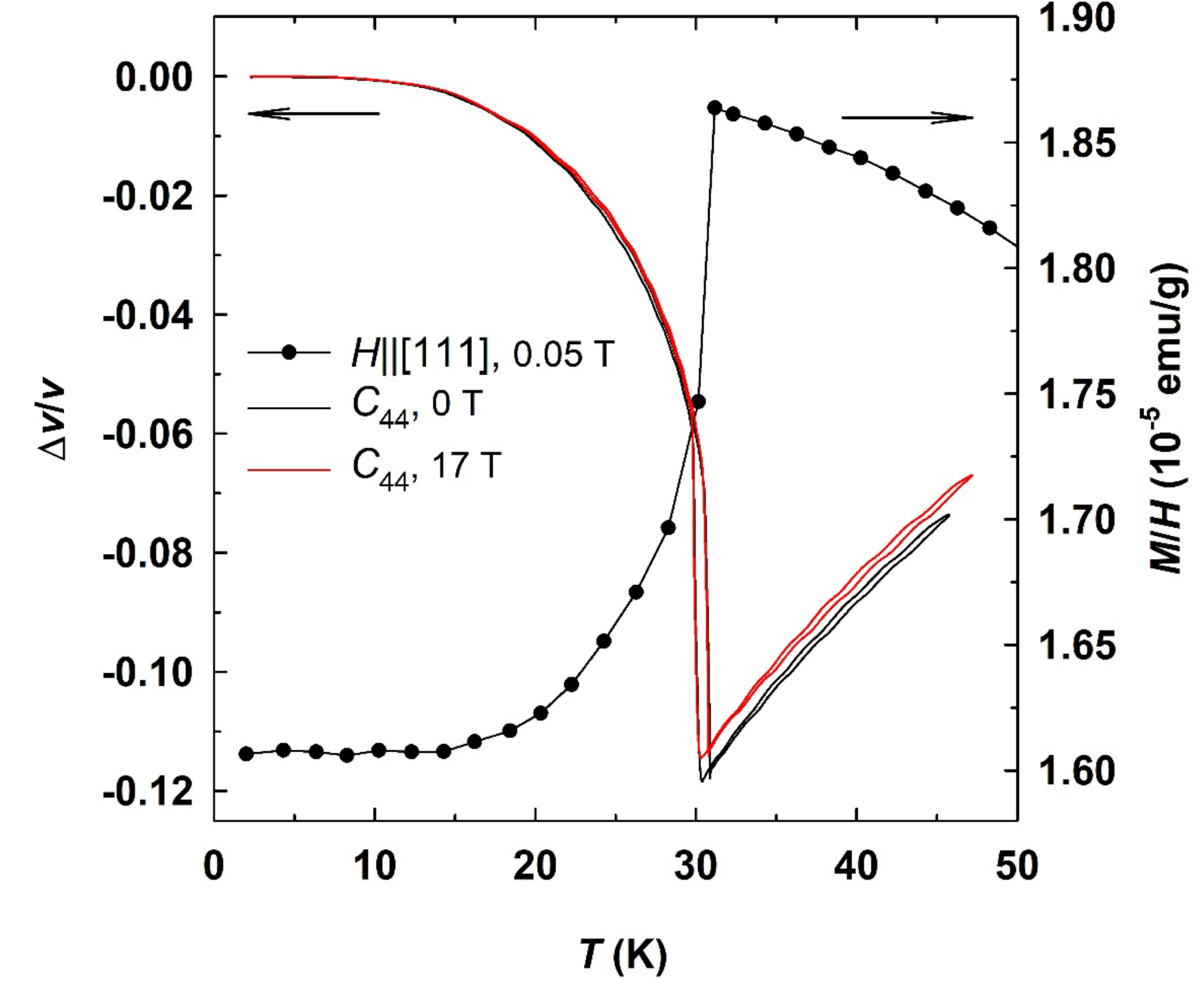}
\caption{Comparison of the temperature dependence of the magnetic susceptibility measured in a 0.05 T field applied along the easy [111] axis of the UO$_2$ single crystal (right scale) with sound velocity, $\Delta v/v$, for the transverse acoustic mode $C_{44}$ in the UO$_2$ single crystal as a function of temperature in 0 and 17 T magnetic fields (left scale). 
}
\label{Fig2}
\end{figure}

The anomalous behavior of $C_{44}$($T$) extends beyond T$_N$ confirming that lattice-spin interactions in UO$_2$ are strong not only in the magnetically ordered state but also well into the paramagnetic region, consistent with Ref. \cite{brandt67}. While the narrow temperature hysteresis of $\Delta v/v$ observed at T$_N$ in UO$_2$ (Fig. \ref{Fig2}) is related to the first-order nature of the transition, it remains unaffected by an applied magnetic field of 17 T. The discrepancy between the $C_{44}$($T$) curves for 0 and 17 T fields in the paramagnetic region above T$_N$ could possibly be attributed to uncorrelated dynamic Jahn-Teller distortion with a wavevector 1\textbf{k}, as reported by Caciuffo \textit{et al.} \cite{caciuffo99}.

\begin{figure}[t]
\centering
\includegraphics[width=1.0\columnwidth]{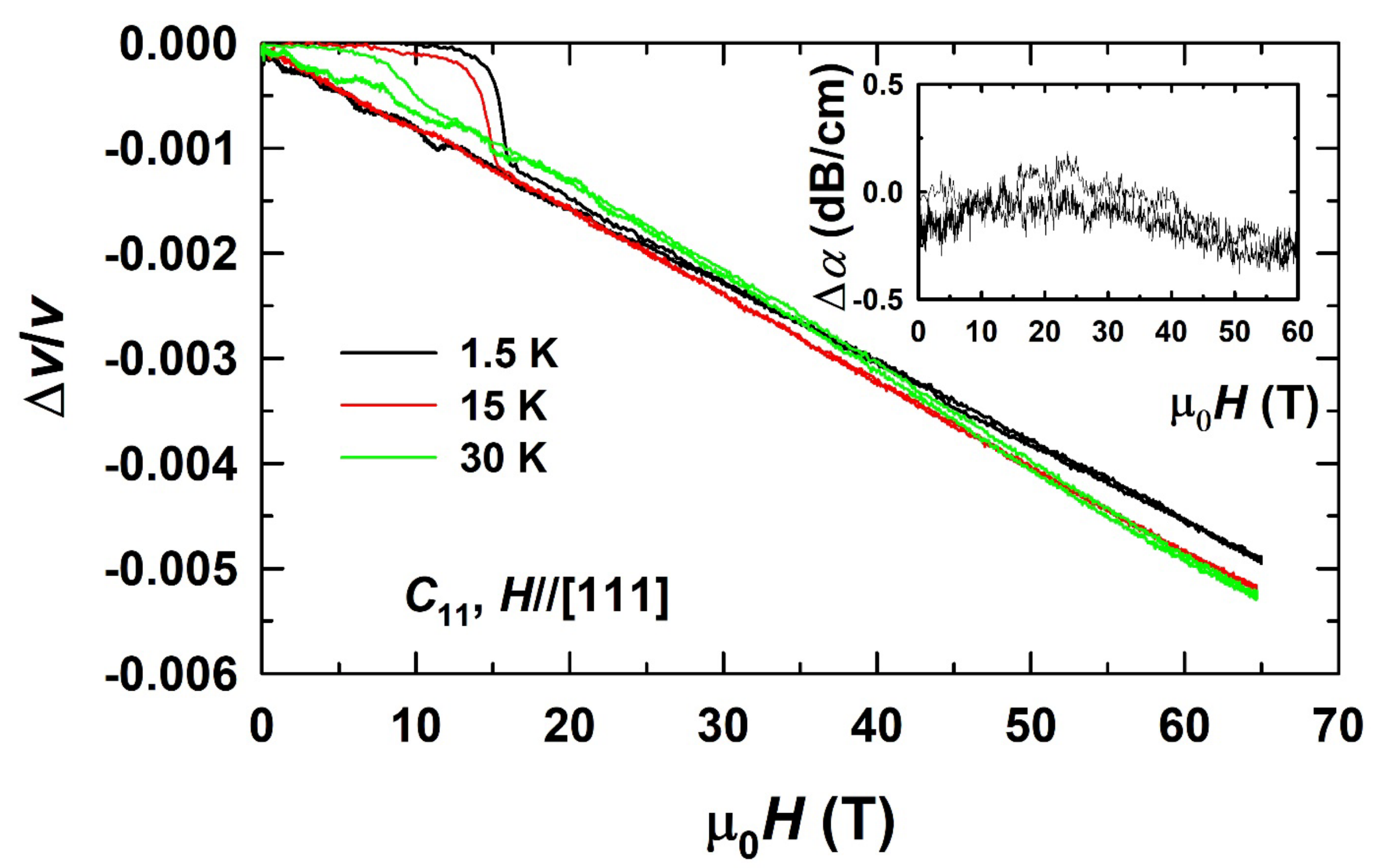}
\caption{Relative sound-velocity change, $\Delta v/v$ for the longitudinal acoustic mode $C_{11}$ in the field applied along the [111] direction of UO$_2$. Inset: the field dependence of sound attenuation $\Delta \alpha$ at 1.5 K.
}
\label{Fig3}
\end{figure}

In the magnetic state at zero magnetic field, there are four distinct antiferromagnetic domains (neglecting time-reversal symmetry) in the UO$_2$ single crystal with $\langle$111$\rangle$-oriented magnetic moments. With the increasing field, these domains are repopulated, leading to the characteristic \textit{butterfly} memory loop observed for fields ±18 T reported by Jaime \textit{et al.} \cite{jaime17}. Our ultrasound measurements reveal significant anomalies in the field dependence of the $C_{11}$ elastic constant along the [111] direction in the ordered phase. $C_{11}$($H$) exhibits a large hysteretic transition in fields below 17 T (see Fig. \ref{Fig3}), which is consistent with the magnetostriction data \cite{jaime17}. Additionally, we observe a narrow hysteresis extending up to $\sim$30 T, suggesting that full domain alignment at the lowest temperatures might occur at fields higher than those previously reported \cite{jaime17,schoenemann21}. Notably, no specific features are observed in the sound attenuation behavior up to 60 T applied along the [111] direction (Fig. \ref{Fig3} (inset)).

The strong piezomagnetic response in UO$_2$ was found to be accompanied by only subtle hysteresis effects in the magnetization data, as reported by Jaime \textit{et al.} \cite{jaime17}. This difference may stem from the distinct responses to the field of the physical mechanisms involved. While the application of a sufficiently high magnetic field along one of the easy [111] directions induces a rhombohedral distortion in UO$_2$ — evidenced by the splitting of the (888) diffraction peak into two \cite{schoenemann21} — the hysteresis in magnetization remains minimal, indicating low energy losses in the process. This could be attributed to the exceptionally stable antiferromagnetic state in UO$_2$, which shows no significant changes even under high (60 T in magnetization \cite{jaime17}) and ultrahigh (150 T in lattice dilation studies \cite{antonio21}) magnetic fields applied along the easy [111] axis. Alternatively, this could result from the gradual and continuous changes in the magnetic configurations occuring in the field, as we propose below.

\begin{figure}
\centering
\includegraphics[width=1.0\columnwidth]{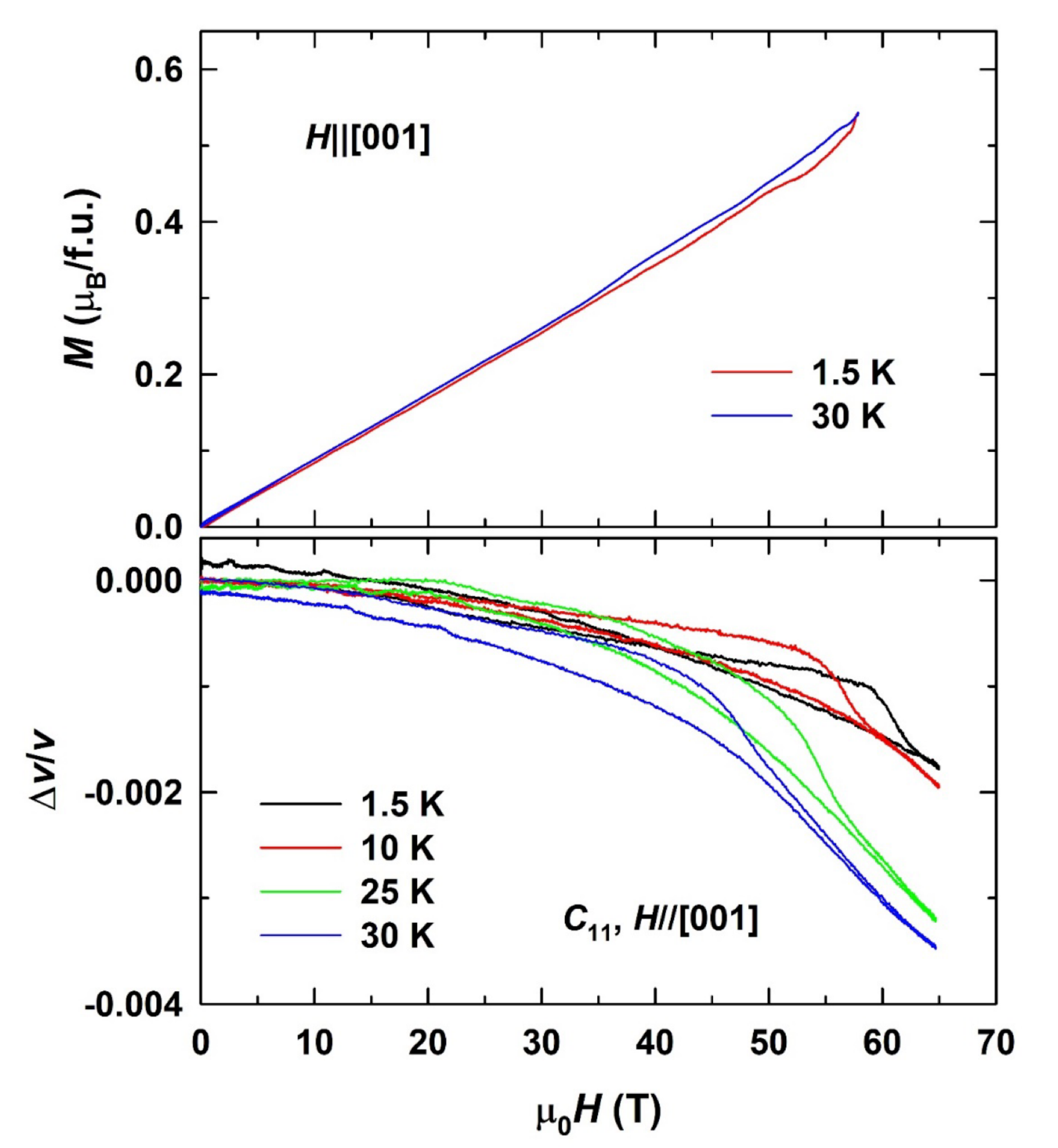}
\caption{(Top) Magnetization of the UO$_2$ single crystal measured along the [001] direction at 1.5 and 30 K. (Bottom) Relative sound-velocity change, $\Delta v/v$, for the longitudinal acoustic mode $C_{11}$ in the field applied along the [001] axis.}
\label{Fig4}
\end{figure}

The most intriguing behavior is displayed by the $C_{11}$ elastic constant with the magnetic field applied along the [001] direction (Fig. \ref{Fig4} (bottom)). We first observe a gradual $C_{11}$ softening accompanied by a broad hysteresis. The softening becomes more pronounced above 50 T, where, as discussed below, our theoretical calculations suggest that the initial 3\textbf{k} magnetic structure is no longer stable under the applied magnetic field. At the same time the high-field magnetization data along the [001] direction in Fig. \ref{Fig4} (top) do not reveal any apparent changes in the magnetic state, i.e. similarly to the magnetization reported for the [111] direction \cite{jaime17}. For both measurement directions [111] (from Ref. \cite{jaime17}) and [001] in Fig. 4 (top), the induced magnetic moment ranges between 0.5 and 0.6 $\mu_B$ per uranium atom, which is still significantly lower than the estimated saturation value of 1.7 $\mu_B$/U.

\begin{figure}[t]
\centering
\includegraphics[width=1.0\columnwidth]{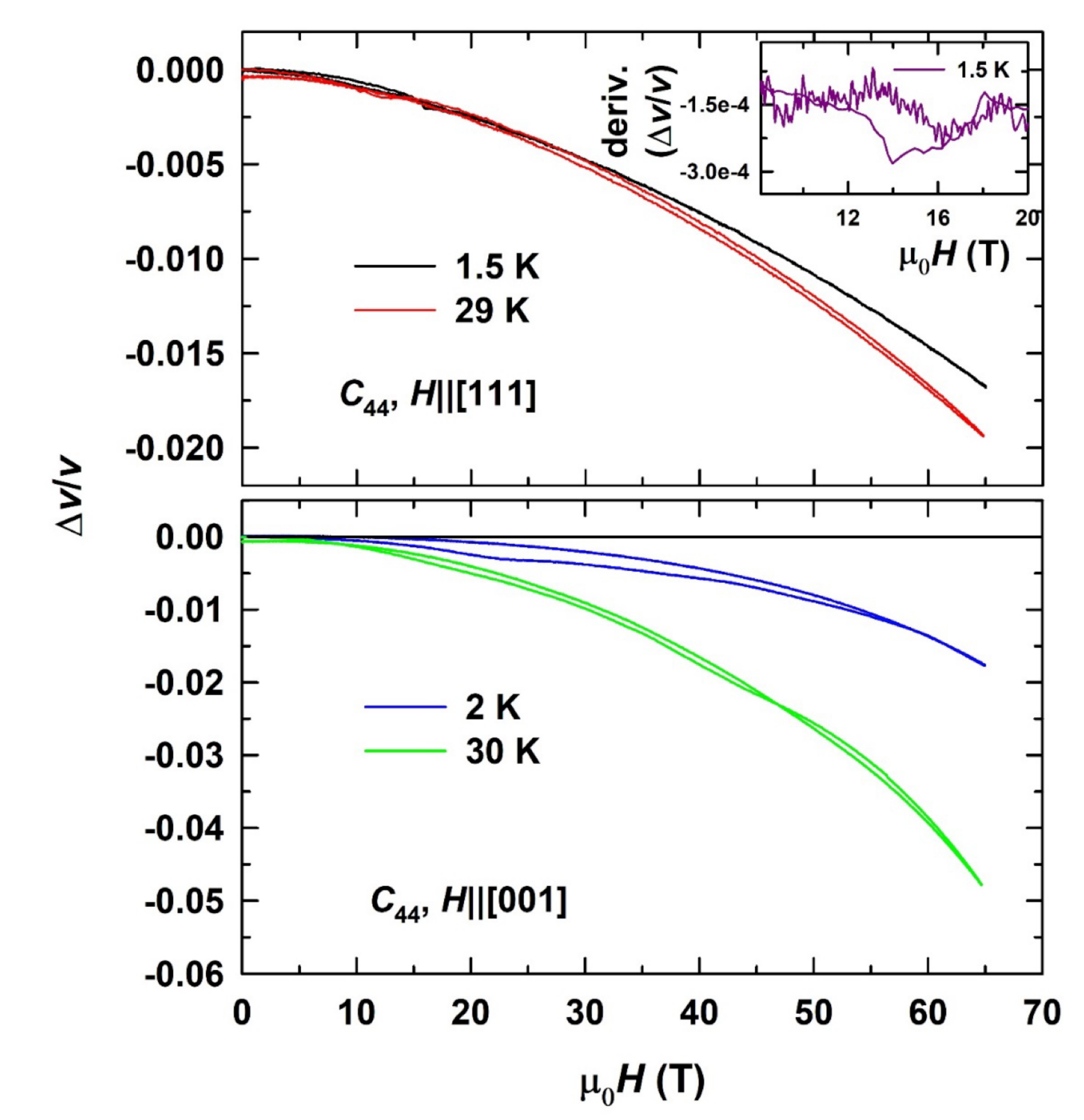}
\caption{(Top) Relative sound-velocity change, $\Delta v/v$, for the transverse acoustic mode $C_{44}$ in the field applied along the [111] direction (top) and [001] direction (bottom) of UO$_2$. Inset shows the derivative of $\Delta v/v$ for the enlarged area below 20 T field. }
\label{Fig5}
\end{figure}

The evolution of the $C_{44}$ elastic constant under a magnetic field applied along the [111] and [001] directions is illustrated in Fig. \ref{Fig5}. $C_{44}$($H$) progressively decreases for both field orientations, indicating a continuous change in the material's stiffness. Notably, the softening observed in $C_{44}$ is an order of magnitude greater than that seen in the $C_{11}$ constant, underscoring its pronounced sensitivity. A small hysteresis associated with piezomagnetic switching is observed in the $\Delta v/v$ derivative for fields below 17 T (see inset in Fig. \ref{Fig5}), while slight hysteresis is noted at the lowest temperatures in fields up to 50 T along the [001] direction.

We demonstrate below that the anomalies in the $C_{11}$ and $C_{44}(H)$ curves likely stem from a change in the magnetic structure symmetry under an applied field along the hard [001] magnetization direction. This assessment is based on our theoretical analysis of the magnetization process in UO$_2$, utilizing \textit{ab initio} calculated super-exchange interactions. As shown in the next section, this analysis predicts a field-induced transition from a 3\textbf{k} to a 2\textbf{k} structure occurring at approximately 50 T in UO$_2$.

\subsection{\textit{Theoretical model and analysis}}
To analyze the possible changes in the magnetic structure of UO$_2$ in applied field, responsible for the anomalous behavior around 50 T, we calculate the magnetization process using the mean-field approximation to the extended Heisenberg model, which includes dipolar and quadrupolar exchange interactions between U$^{4+}$ ions and the Zeeman term \cite{santini09}: 

\begin{equation}
\hat{H} = \sum_{ij} \sum_{KQK'Q'} V^{ij}_{KQ,K'Q'} \hat{O}^{Q}_{K(i)} \hat{O}^{Q'}_{K'(j)} + \sum_i g_J \mu_B \vec{H}_{ext} \vec{\hat{J}}_i
\end{equation}

\noindent
where the summations run over the $i,j$  U-lattice sites. $\vec{H}_{ext}$ is the applied external magnetic field, $\mu_B$ is the Bohr magneton, $g_J$ =4\/5 is the Land\'e factor for the $^{3}H_{4}$ ground state multiplet of the U$^{4+}$ ion, and $\vec{\hat{J}}_i$ is an angular momentum operator on the $i^{th}$ ionic site. The inter-site exchange interactions between spherical tensor (or normalized Stevens) operators is defined within the S = 1 pseudospin basis of the lowest crystal field $\Gamma_5$ triplet of the $^{3}H_{4}$ multiplet. Within the $\Gamma_5$ triplet, the angular momentum operators are related to the corresponding spherical tensors as $J_Q$ = $2\sqrt{2} \hat{O}^{Q}_{1}$ , where $Q$ = $x,y,z$. In the case of correlated insulators, the two-site interaction (SEI) parameters $V^{ij}_{KQ,K'Q'}$ are due to superexchange interactions. In this study we take SEI parameters from our earlier \textit{ab initio} calculations \cite{pourovskii19} where they were derived for UO$_2$ on the basis of a first-principles magnetic force theorem \cite{pourovskii16}, in the framework of a density-functional theory+many-body Hubbard-I method \cite{hubbard63,lichtenstein98}. The experiment \cite{nakotte10} and these calculations \cite{pourovskii19} also have shown that the CEF splitting in UO$_2$ is rather large ($\sim$150 - 180 meV), and thus the model based on the isolated spherically symmetrical $\Gamma_5$ triplet should serve as a very good approximation for the modeling the magnetization process in UO$_2$ in up to hundreds tesla fields. 

\begin{figure}[h]
\centering
\includegraphics[width=1.0\columnwidth]{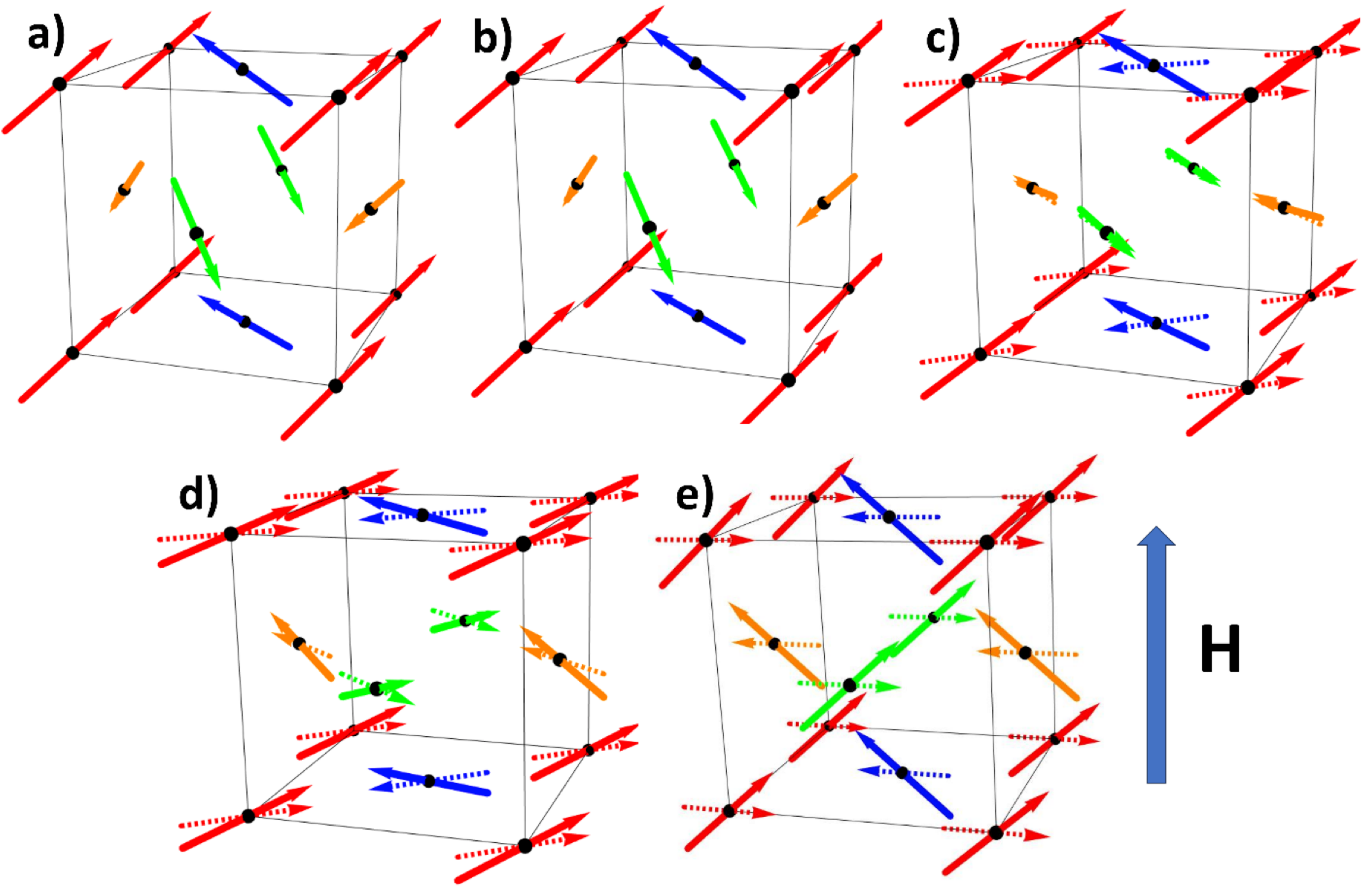}
\caption{Magnetic structures at various fields denoted with the vertical arrow and applied along the [001] direction: a) 0 T, b) 20 T, c) 40 T, d) 60 T, e) 120 T. The four magnetic sublattices of the 3\textbf{k} structure are indicated by color. The solid and dashed arrows are full moments and their projection to the ($x,y$) plane, respectively.}
\label{Fig6}
\end{figure}

In Fig. \ref{Fig6}, we present the calculated magnetic structures of UO$_2$ under various fields applied along the [001] direction (vertical axis in the figure). Significant symmetry changes occur around the 50 T field. The structures below this field (panels a-c in Fig. \ref{Fig6}) maintain the symmetry of the original 3\textbf{k} ground state. However, above 50 T (Fig. \ref{Fig6}d), the $z$-components of the U sublattices' magnetizations align along the applied field direction, while the in-plane components (indicated by dashed arrows in panels c) and d) form a 2\textbf{k} structure. Moreover, our simulations also predict a second transition from the 2\textbf{k} structure to a 1\textbf{k} to occur at $\sim$104 T. The predicted 1\textbf{k} structure is shown in Fig. \ref{Fig6}e. We note that both transitions take place through a smooth rotation of the moments, without any discontinuity in the magnetization. First, the moments rotate to eliminate the AF component along the field thus forming \textbf{2}k structure with the moments lying in {110} planes. Subsequent increase of the field from 50 to 104 T leads to the moments rotation from the {110} to {100} planes, thus transforming the AF structure into 1\textbf{k}.

\begin{figure}
\centering
\includegraphics[width=1.0\columnwidth]{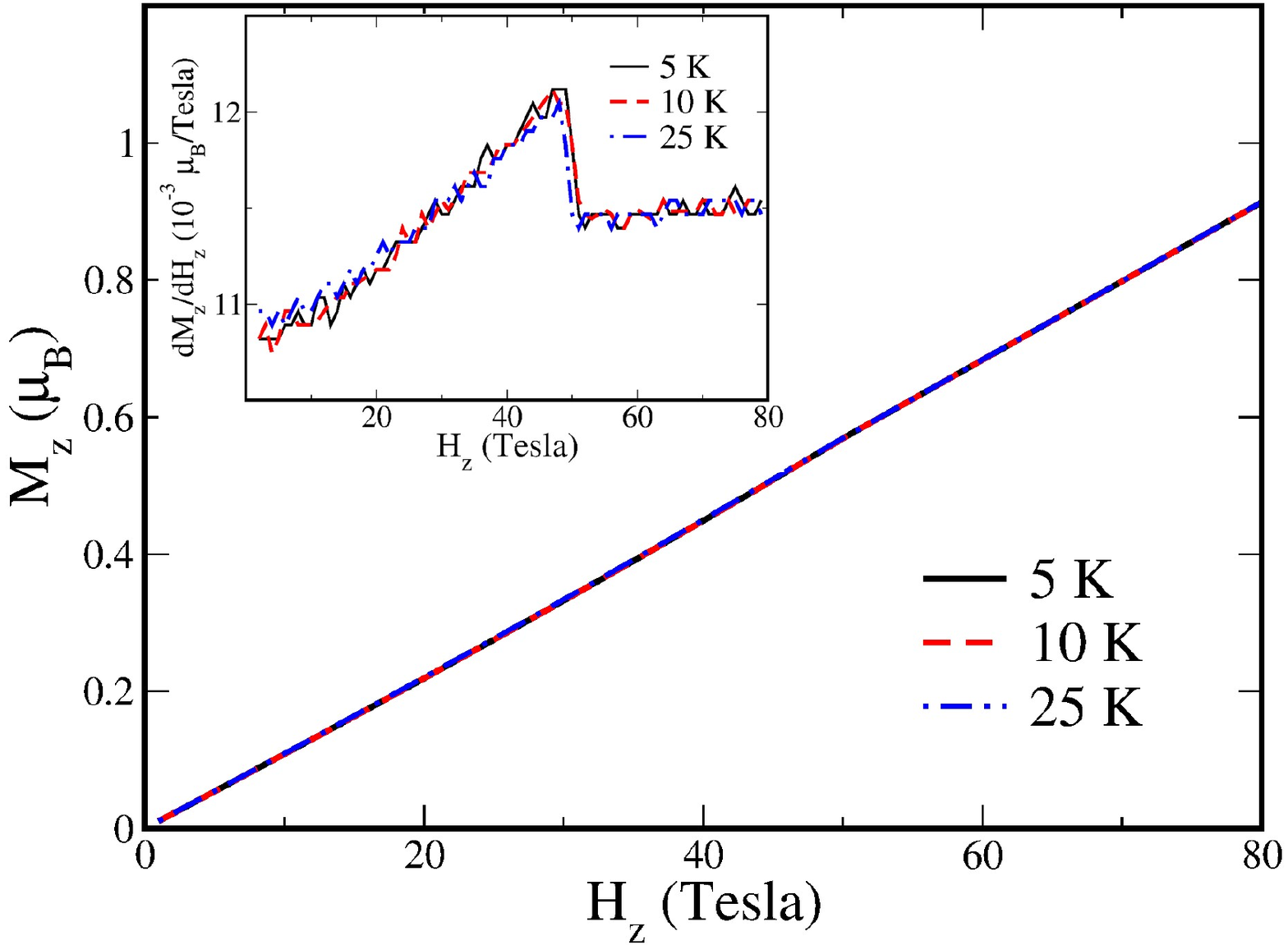}
\caption{The calculated magnetization curves and the corresponding derivatives $dM_z$/$dH_z$ vs. $H_z$ (inset) at various temperatures.
}
\label{Fig7}
\end{figure}

\begin{figure}
\centering
\includegraphics[width=1.0\columnwidth]{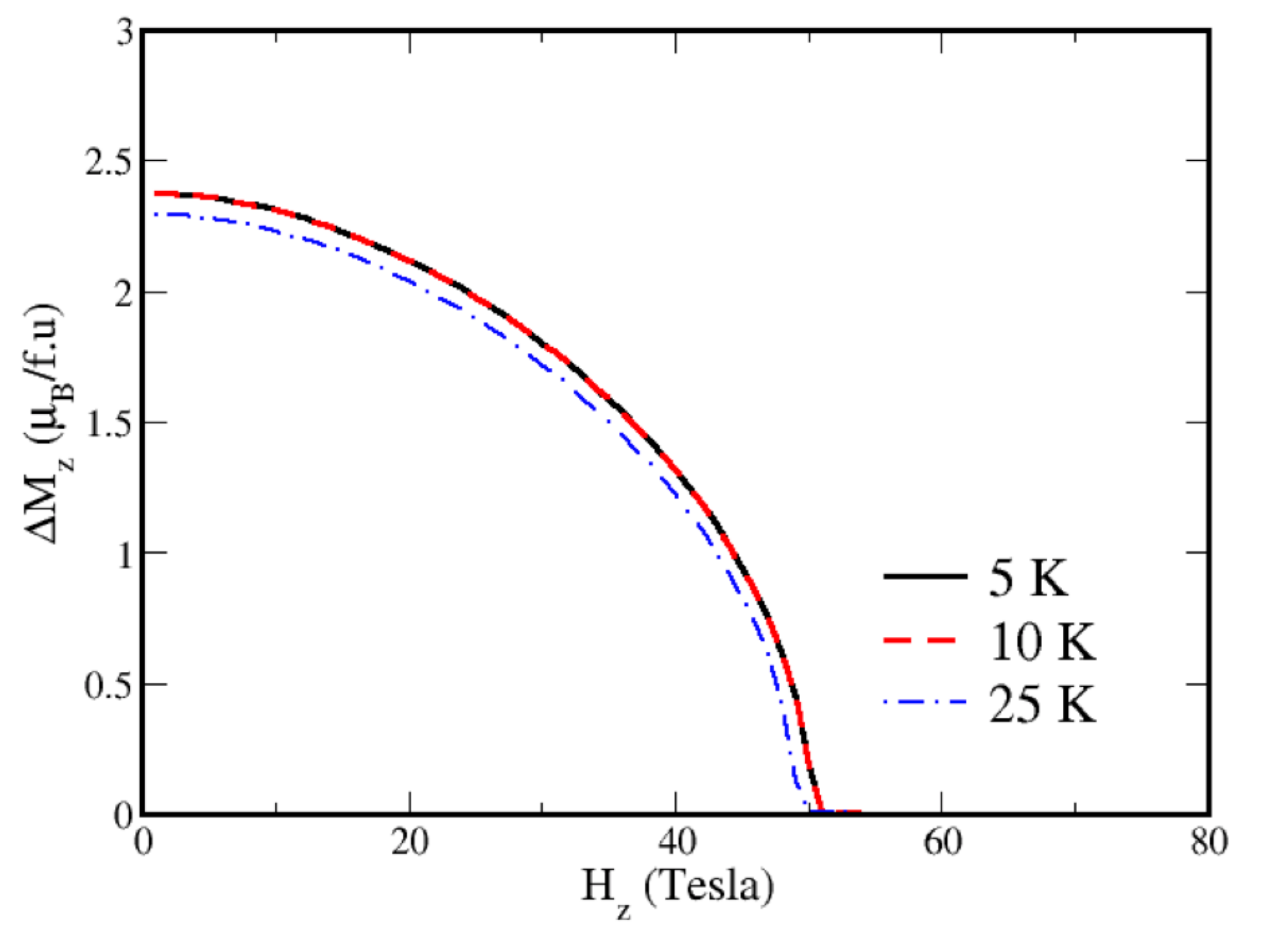}
\caption{The difference in $M_z$ vs. $H_z$ between the two sites in 3\textbf{k} structure at various temperatures.
}
\label{Fig8}
\end{figure}

Let us now focus on the transition associated with a change in the magnetic structure from 3\textbf{k} to 2\textbf{k} at around 50 T, which is within the range of fields in our experiment. Although this transition is not evident in the magnetization curves (Figs. \ref{Fig4} (top) and \ref{Fig7}), it is clearly indicated in the corresponding derivative $dM_z/dH_z$ vs. $H_z$ curve (inset in Fig. \ref{Fig7}). Beyond 50 T, the structure remains antiferromagnetic in the ($x,y$) plane but becomes ferromagnetic in the $z$ direction. This difference is evident in the different $M_z$ vs. $H_z$ between the two sites, where $M_z$ components are opposite in the pure 3\textbf{k} structure (Fig. \ref{Fig8}). This difference disappears at 50 T, indicating the transition. The transition is continuous, and the emergence of the new symmetry element can be attributed to the changes in the $\textbf{M}$ $\|$ $\textbf{H(001)}$  components of the sublattices' magnetizations.

\begin{figure}[h]
\centering
\includegraphics[width=1.0\columnwidth]{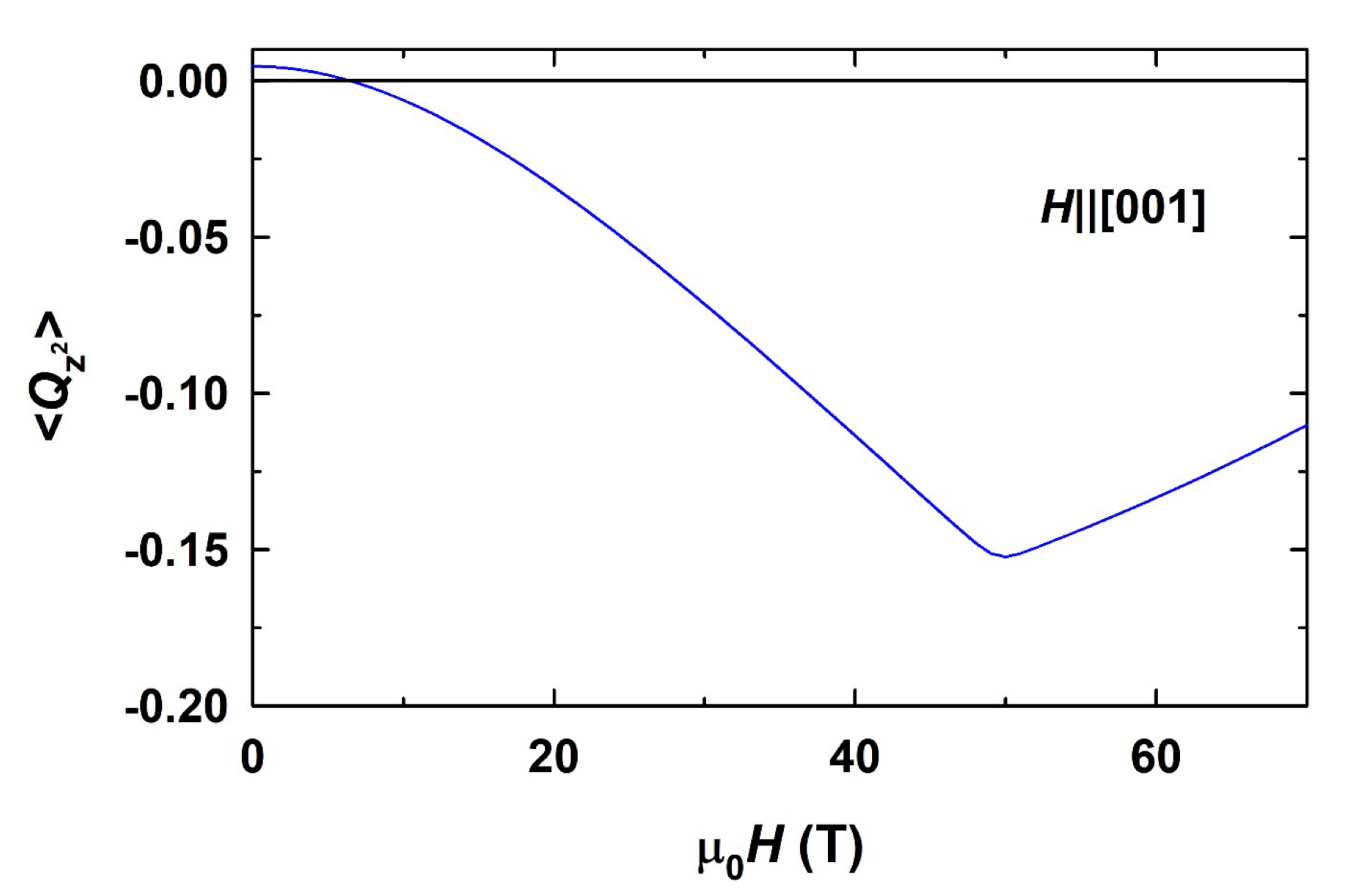}
\caption{The quadrupole moment $\langle Q_{z^{2}}\rangle$ averaged over the 3\textbf{k} magnetic cell vs. magnetic field applied along the [001] direction.
}
\label{Fig9}
\end{figure}

The uranium on-site quadrupole moments in UO$_2$ are rigidly linked to the on-site magnetic moments. Hence, the discontinuity in $dM_z/dH_z$ vs. $H_z$ is expected to induce the corresponding peculiarity in the evolution of quadrupole moments. Indeed, our simulations also predict the expectation value of the $z^2$ quadrupole $\langle O_{z^{2}} \rangle$ $\propto$
$\langle 3J_{z}^{2}-J(J+1) \rangle$ to exhibit a first-derivative discontinuity at the transition point of 50 T, as shown in Fig. \ref{Fig9}.

The quadrupole structure and elastic constants are related by a second-order anisotropic magnetoelastic coupling, as discussed in Ref. \cite{tremolet78}, where a general form of this coupling for cubic crystals is derived (see Eq. 30 therein). In particular, their expression contains the term $B_{zz} O_{z^2} \epsilon_{zz}^2$, where $\epsilon_{zz}$ is the diagonal symmetric component of the strain tensor. One can see that the first-derivative discontinuity in $\langle O_{z^2} \rangle$ vs. H will lead to a corresponding discontinuity in the $C_{11}$ elastic constant. Therefore, the predicted change in the quadrupolar structure qualitatively explains the observed peculiarity in the sound velocity. We notice, however, that the second-order magnetoelastic coupling expression in \cite{tremolet78} also contains the coupling $B_{xy} O_{z^2} \epsilon_{xy}^2$  between the off-diagonal stain tensor component $ \epsilon_{xy}$ and the same quadrupole. This coupling is expected to cause the same type of peculiarity in $C_{44}$. The magnetoelastic coupling constants for the two cases are though different. Hence, our results suggest that the anomaly in $C_{44}$ exists, but it likely falls below the detection limit of our experiment. We notice that the relative change of $C_{44}$ in field (about 2\% between 0 and 60 T, at the lowest temperature) is much larger than that of $C_{11}$ (about 0.15\% between 0 and 60 T, at the lowest temperature). Hence, an anomaly in $C_{44}$ of the same relative magnitude as observed in $C_{11}$ would be hardly visible. Direct \textit{ab initio} calculations of high-rank magnetoelastic coupling constants would be highly desirable to clarify this issue. Unfortunately, in a system as complex as UO$_2$ they are not feasible at present.

\section{Conclusions}
We have carried out ultrasound measurements for the UO$_2$ single crystal in high magnetic fields and found that the $C_{11}$ elastic constant exhibits an anomaly under fields $\sim$50 tesla applied in the hard [001] direction. Using theoretical simulations on the basis of an \textit{ab initio} superexchange Hamiltonian, we identify the origin of this anomaly as a transition from the 3\textbf{k} to the 2\textbf{k} structure occurring right above 50 T. Furthermore, we anticipate a subsequent transition to a 1\textbf{k} antiferromagnetic structure at around 104 T. Both transitions are due to continuous rotations of the moments direction leading to disappearance of one of the antiferromagnetic components. Quadrupole moments of U are found to exhibit an anomaly vs. field at the 3\textbf{k} $\rightarrow$ 2\textbf{k} transition, which can explain the observed anomaly in the $C_{11}$ elastic constant. A 3\textbf{k} $\rightarrow$ 2\textbf{k} transition has also been speculated to occur in the surface magnetism of UO$_2$ \cite{langridge14} employing grazing incidence x-ray resonant magnetic scattering. Of course, in none of these examples has the transition been unambiguously proven. This would require careful neutron studies, though the challenge lies in the need for relatively large samples, while such high magnetic fields typically maintain uniformity only over a small volume.

Our present results complement previous observations of a 3\textbf{k} $\rightarrow$ 1\textbf{k} transition induced by a tetragonal strain in UO$_2$ thin films \cite{tereshina24}. The present study together with Ref. \cite{tereshina24} thus shows that the 1\textbf{k} and 2\textbf{k} AF orders, which compete on the frustrated UO$_2$ lattice structure with the 3\textbf{k} ground state, can be stabilized by various external stimuli applied along the [001] direction. Quantitative experimental studies of strain along other crystallographic directions are desirable to deepen our understanding of the interplay between magnetic order, strain, and applied field in UO$_2$. Moreover, our present theoretical analysis is based on purely electronic superexchange model; the role of electron-lattice coupling, which is likely significant in UO$_2$ \cite{jaime17,paolasini21}, in the relative stability of various magnetic orders also needs to be clarified. 
	
\section{Acknowledgments}
We would like to thank Prof. A.V. Andreev for his assistance in making the samples ready for measurements, including their cutting and orientation, and to Prof. G. H. Lander for valuable comments. We acknowledge the support of Czech Science Foundation under the grant no. 22-19416S. We acknowledge support of the HLD at HZDR, a member of the European Magnetic Field Laboratory (EMFL). Physical properties measurements were performed in the Materials Growth and Measurement Laboratory (http://mgml.eu/) supported within the program of Czech Research Infrastructures (Project No. LM2023065). L.V.P. is thankful to the CPHT computer team for support.

\bibliography{UltrasoundUO2}

\end{document}